\documentclass{aa}    
\usepackage{graphicx}
\usepackage{pdflscape}
\usepackage{subfigure}

\begin{document} 

\def\etal{{\rm et al. }}
\def\mpc{{\  h^{-1} \rm Mpc}}
\def\kpc{{\ h^{-1} \ \rm kpc}}
\newcommand\kms{{\rm km~s$^{-1}$}}


   \title{Analysis of interacting and isolated quasars}

   \author{L. Donoso,\inst{1}
           M. V. Alonso,\inst{1,2}
           D. Garc\'ia Lambas,\inst{1,2}
           G. Coldwell,\inst{3,4}
           E. O. Schmidt,\inst{1} and 
           G. A. Oio\inst{1}
          }

   \institute{Instituto de Astronom\'ia Te\'orica y Experimental, (IATE-CONICET), Laprida 854, C\'ordoba, Argentina
         \and
      Observatorio Astron\'omico de C\'ordoba, Universidad Nacional de C\'ordoba, C\'ordoba, Argentina
                  \and
        Consejo Nacional de Investigaciones Cient\'ificas y T\'ecnicas (CONICET), Argentina
         \and
            Departamento de Geof\'isica y Astronom\'ia. Facultad de Ciencias Exactas, F\'isicas y Naturales, Universidad Nacional de San Juan,
                San Juan, Argentina
                       }

\date{Received; accepted}

 
  \abstract
   {}
{The main goal of this study was to determine the effects on equivalent
widths (EWs) of some spectral lines produced in the quasars by the presence of
surrounding galaxies. To carry this out, a sample of 4663 quasars (QSOs) in the 
redshift range of 0.20 to 0.40 from the Sloan Digital Sky Survey--Data 
Release 7 was analyzed. }
{Three
QSO sub-samples were defined, taking into account
the projected separations
and radial velocity differences with neighboring galaxies. In this way,
we utilized two sub-samples of QSOs with strong and weak galaxy interactions, with 
projected separations smaller than 70 kpc, and between 70 and 140 kpc,
respectively, and with radial 
velocity differences less than $5000~$\kms.  These sub-samples were
compared with isolated QSOs defined as having greater 
projected separations and radial velocity 
differences to the galaxies. } 
   {From a statistical study of the EWs of relevant
     spectral lines in the QSOs, we show an increment of the EWs of about 20\% 
     in the [OIII]$\lambda\lambda4959,5007$ lines and 7\% in H$_{\alpha}$
     for QSOs
     with stronger galaxy interactions relative to
the isolated QSOs.  These increments were also observed restricting the
sub-samples to velocity differences of $3000~$\kms.
These results indicate that some line EWs of QSOs could be marginally influenced
by the environment and that they are not
affected by the emission of the host galaxy, which was estimated to be around 
10\% of the total emission.

Furthermore, in order to gain a better understanding of the origin of the 
H$_{\alpha}$ emission line, we performed broad and narrow line decomposition
in 100 QSOs  in the restricted Sint sub-sample and also 100 randomly selected QSOs in the Iso sub-sample.
 When these QSOs were compared, the narrow component 
remained constant  whereas the broad component was incremented.
Our results, which   reveal slight differences in EWs of some emission lines, suggest that galaxy interactions with QSOs may
affect the
QSO activity.
   }
{}


   \titlerunning{Analysis of interacting and isolated quasars}
   \authorrunning{Donoso et al.}         

   \maketitle
%

\section{Introduction}

   Active galactic nuclei (AGN) typically emit large amounts of radiation 
at all frequencies, with the observed spectra being composed of a non-thermal 
continuum with emission and absorption lines arising from small central 
regions, only a few parsecs across (Osterbrock 1989, Carswell et al. 1976). 
The AGNs are
powered by gas, dust, and stars in a disk when falling into a super-massive
black hole (Zel'dovich \& Novikov 1965; Lynden-Bell 1969). To model
AGNs, 
 the existence of different regions is taken into account, such as
the broad (BLR) and narrow line (NLR) regions, and the accretion disk and jets 
(Antonucci 1993; Urry \& Padovani 1995).

Quasars (QSOs) are 
the most luminous AGNs, in which the optical images are dominated by a blue,
very luminous, and variable unresolved nucleus (Schmidt 1969).  
The observed spectrum is the result of the integrated spectrum of both the
host galaxy and the nucleus associated with the AGN.
An important 
characteristic of the spectra is the presence of strong
emission lines that can be used to infer both the physical 
conditions of the
emitting gas 
and the properties of the ionization sources (Strittmatter 
\& Williams 1976).  In this sense, the broad lines, such as the Balmer series, 
are produced in the innermost 
regions (BLR, a few parsecs from the nucleus), that surround the
accretion disk where the gas densities and velocities are high.
These are the brightest recombination lines
in the optical spectra of AGNs, and normally have complex line
profiles that often exhibit both broad and narrow components 
emitted from physically distinct regions.
In contrast, the narrow lines have their origin
in more extended regions  of about 50 pc in size, or NLRs, 
which are 
characterized by lower gas densities and velocities. The most important
lines that are thought to be emitted within the NLRs of the AGNs are [OII], 
[OIII], [NeIII] and [NeV] (Osterbrock 1989).
Heckman
et al. (2004) and Kauffmann et al.
(2003) have also suggested that the luminosity of the line [OIII]$\lambda5007$ is
a good
tracer of AGN activity. This line is the strongest narrow emission 
line in 
the optical part of the spectrum with low contamination from the 
contribution of star formation in the host galaxy.

The intense brightness of QSOs puts serious constraints on the study of the 
host 
galaxies.  However, at lower 
redshifts, the use of the Hubble Space Telescope has made possible the
host galaxy observations, with interesting details in their morphologies.
Bahcall, Kirhakos \& Schneider (1997) studied luminous QSOs in different environments, and reported
normal ellipticals and spirals and also strongly disturbed or 
interacting systems at z $<$ 0.3.  Floyd et al. (2004) found that QSOs with nuclear 
luminosities
M$_V < $ -24 mag and z $<$ 0.4 are massive bulge-dominated galaxies, thereby
 confirming the main results of
Dunlop et al. (2003).  At even lower redshifts (z $<$ 0.2), 
Jahnke, Kuhlbrodt \& Wisotzki (2004) 
observed the presence of symmetric disk and elliptical host galaxies,
suggesting that minor mergers 
or gas accretion are responsible for galaxy activity. More recently, 
Falomo et al. (2014) studied 
galaxy hosts in a QSO sample extracted from the Sloan Digital Sky Survey
Seven Data Release (Abazajian et al. 2009, hereafter SDSS--DR7) Stripe 82 
 at z $<$ 0.5, and reported complex morphologies with bulge- and disk-dominated 
galaxies.  

Hydrodynamical simulations of galaxy mergers
(Di Matteo, Springel \& Hernquist 2005)
showed that two different phenomena may be generated, namely the
production of intense star formation in the host galaxy and a fast
increase in the accretion rate of the black hole.
These interactions also have the potential to transport material 
to the central regions, producing AGN activity (Jogee 2006).
Moreover, the proximity of the companion galaxy could affect the mass 
distribution, leading to gas inflows enhancing star formation activity when 
compared with isolated systems.  Based on these simulations, we were
interested in looking at 
QSO interactions with nearby galaxies to try to determine whether this 
phenomenon can 
generate changes in the 
QSO spectrum when compared with isolated QSOs.  These changes in, for example,
equivalent widths (EWs)  can provide
information about the physical conditions, and the distribution and dynamics of
the emitting gas.  Thus, a statistical analysis
of EW spectral lines can reveal the influence of 
strong and weak galaxy interactions with QSOs.

The paper is organized as follows: 
in $\S$2 we discuss the QSO sample 
and the available photometric and spectroscopic data for QSOs and neighboring
galaxies. Based on typical projected distances and radial velocity differences,
we also present our criteria for defining the
three sub-samples used to analyze QSO interactions with galaxies.
In $\S$3 we discuss the spectral analysis including the statistical analysis
of the EWs and uncertainties in relevant  
spectral 
lines together with the spectral decomposition. Finally, in $\S$4
a summary is presented with the main results of this 
study.  For all cosmology-dependent calculations, we have assumed 
$\Omega_\Lambda$ = 0.7, $\Omega_m $ = 0.3 and H$_0$ = 70 \kms.

\section{Data and samples}

%

In order to study the effects in the spectra of QSOs produced by surrounding
galaxies, we need to define the QSO sample and the 
characterization of the QSO environments.  

\subsection{The quasars}

We have selected QSOs from the SDSS Quasar Catalog--DR7
(SDSSQ--DR7, Schneider et al. 2010), which contains 
spectroscopically confirmed QSOs fainter than $i \approx$ 15.0 mag and having 
at least one broad emission line in the spectrum with full width at half maximum (FWHM) larger than 
1000 \kms~or with complex absorption features.   The QSOs have 
{\it ugriz}~magnitudes with
typical uncertainties of 0.03 mag, and have reliable spectroscopic 
redshifts in the range  0.065 to 5.460 with uncertainties of about 
3$\times$10$^{-4}$ that roughly corresponds to 90 \kms~(Stoughton et al. 2002).
The QSO sample used in this study was defined in the redshift range 
0.20 $ < z \leq $ 0.40 as the only restriction, comprising 4663 objects.
This redshift range is adequate to perform the analysis of the QSO interactions
with galaxies.  The lower limit assures
a significant number of QSOs with statistical significance, and the upper
limit allows us to find galaxies around the QSOs within the completeness levels
of the SDSS.  Although there are new releases of the quasar catalog, DR10 (P\^aris et al. 2014) and
DR12 (P\^aris et al. 2017), these adopt different
procedures to define QSOs since their aim is to study those QSOs at higher
redshifts, 
z $>$ 2.15. The SDSSQ--DR7 is the catalog with
the most objects in our redshift range, (see Fig. 4 of P\^aris et al. 2017) and
therefore better suited for this study.

\begin{figure*}
\centering
\includegraphics[width=70mm,height=70mm]{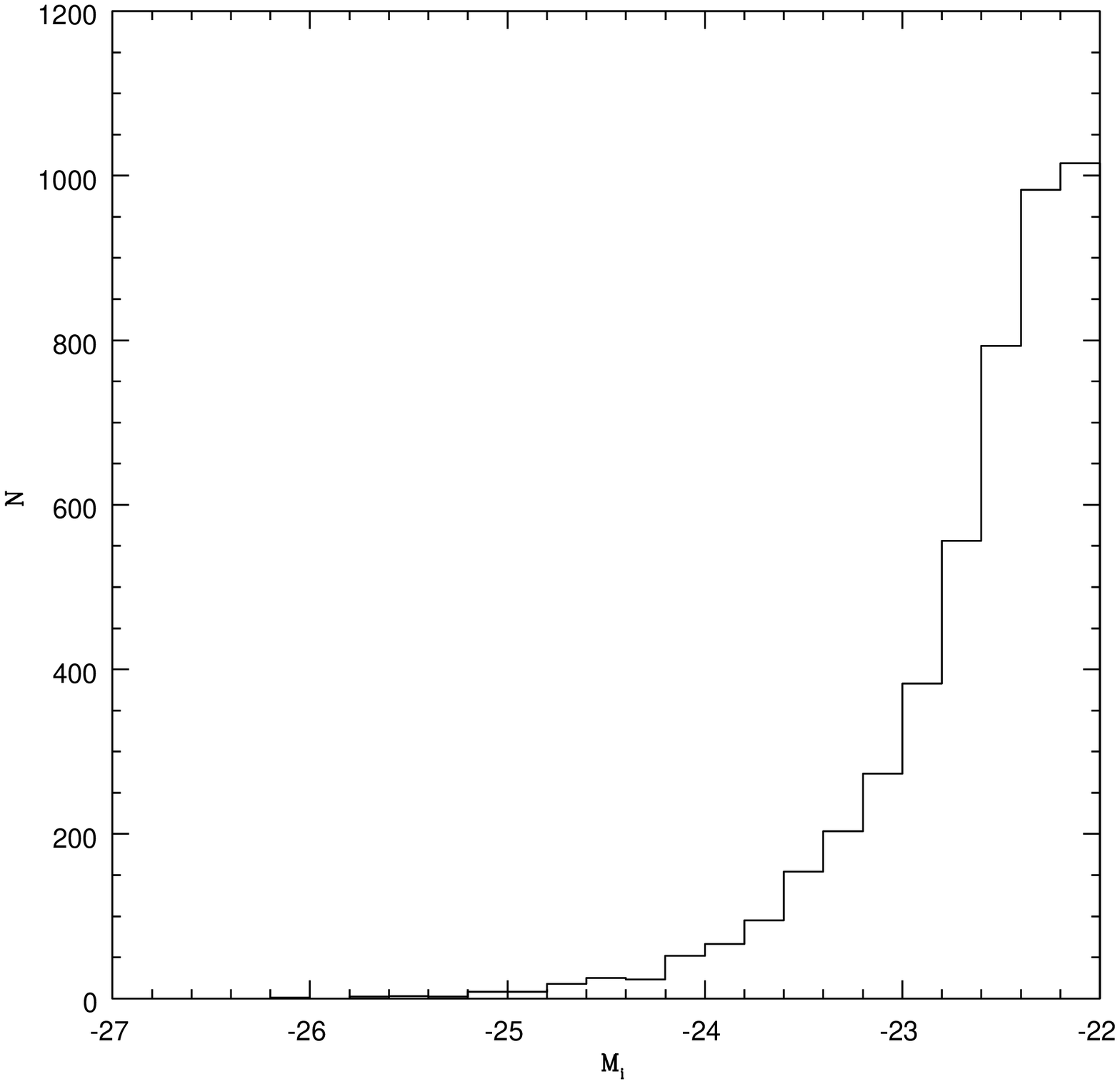}
\includegraphics[width=70mm,height=70mm]{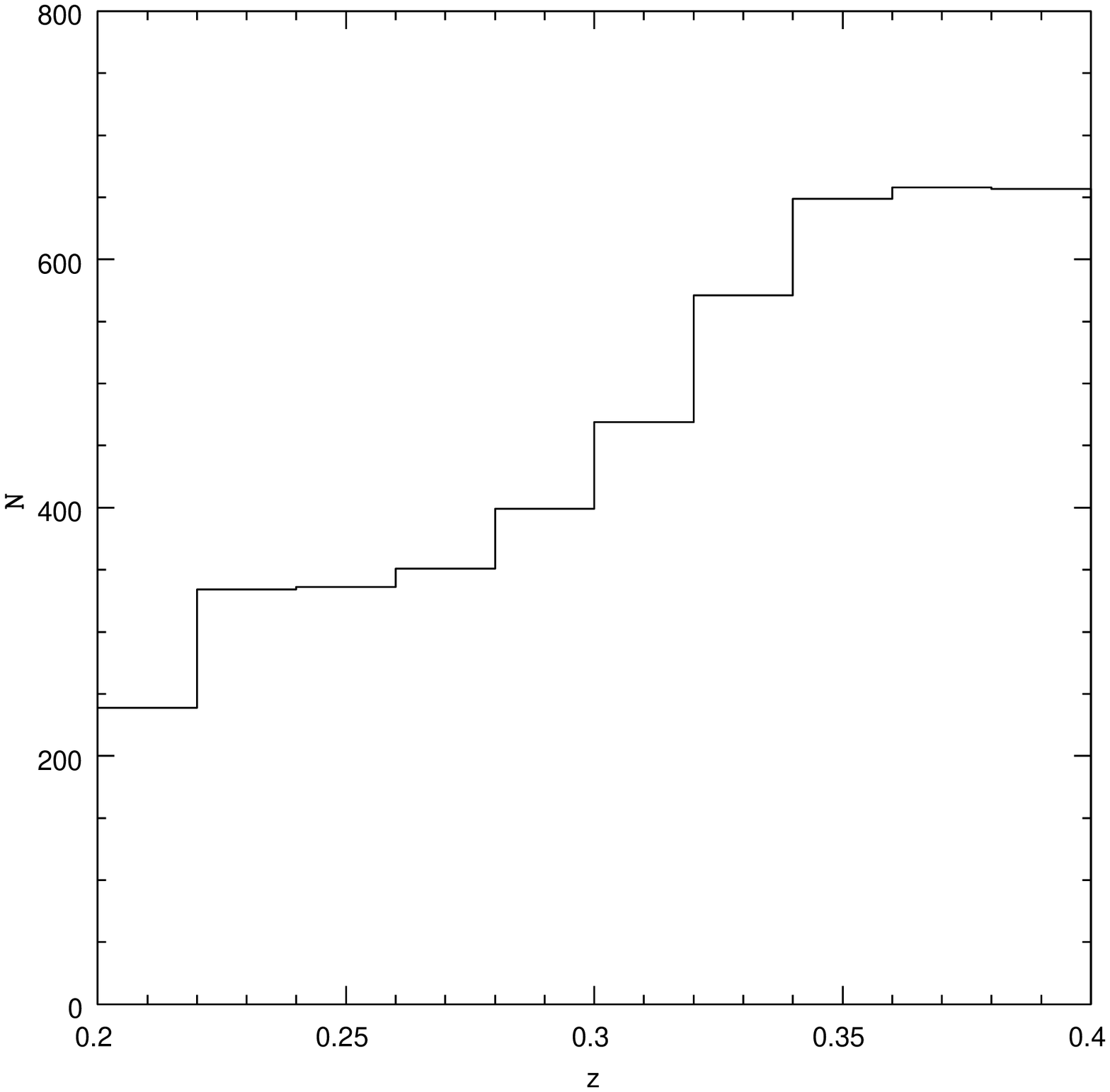}
\caption{The absolute magnitude and redshift distributions of the QSO sample.}
\label{histoM}
\end{figure*}

Figure~\ref{histoM} shows the distributions of absolute magnitudes in the
{\it i} passband and redshifts of the QSO sample.  The  absolute
  magnitudes extend between -26.18 and -22.00 mag in the redshift range from
  0.2 to 0.4. 

   \subsection{Selection criteria}

   In order to establish the QSO interaction with neighboring galaxies, we used
   the projected separation, r$_p$ to the QSO 
   and the relative radial velocity, defined as $\Delta V = c (z_Q - z_g) $,
   where $z_Q$ and $z_g$ are the QSO and galaxy redshift estimates,
   respectively.  In this sense,
   we define the  interaction QSO--galaxy as strong using
   tight  r$_p$ and $\Delta V$ and
   weak, with moderate values.  Also, isolated QSOs were defined
   as being those without neighboring
   galaxies.
   A similar procedure was adopted by Lambas et al. (2003) in a sample of
   galaxy pairs to analyze the interaction
   effects on galaxy star formation rates.

   In order to define the three QSO sub-samples, we used galaxy data 
   obtained from the Galaxy 
   table of the CasJobs database from the SDSS--DR7 (Abazajian
   et al. 2009), which contains the photometric parameters 
of the objects classified as galaxies.  This survey has 
photometric data in the {\it ugriz}~passbands with an r$^\prime$ limiting 
magnitude of 22.20 mag representing a 95\% photometric completeness for 
point sources 
(Abazajian et al. 2004). The spectroscopic information has 
 wavelength coverage in the range 3800-9200 \AA~with a spectral resolution of 
$\approx $ 2000, and the spectroscopic catalog 
is complete with a Petrosian r$^\prime$ magnitude (Petrosian 1976) 
of 17.77 mag (Strauss et al. 2002). 

Most of the galaxies
have redshift information only from photometric estimates.  They
are based, for example, on empirical methods (e.g., Connolly et al. 1995)
that use an artificial Neural  Network and the ANNz code 
developed by Collister \& Lahav (2004).
These methods require a priori redshift
information 
obtained through a training 
set of galaxies with both photometric and precise spectroscopic redshifts.
O' Mill et al. (2011) obtained photometric redshifts and k-corrections for
galaxies with r$^\prime$ = 21.5 mag of the SDSS--DR7 using the
five {\it ugriz}~passbands.  The limiting magnitude guarantees a reasonable
photometric redshift quality and a good separation 
between galaxies and stellar objects (Stoughton et al. 2002,
Scranton et al. 2002).  Their photometric redshift estimates
are based on the empirical methods using the ANNz code and their 
Fig. 1 shows the comparison with the SDSS--DR7 spectroscopic redshfits.
The agreement is excellent, especially in
the range of our study, 0.2 $ < z \leq $ 0.40, with uncertainties of
$\sigma_{phot} \sim 0.0227$. We used O' Mill et al. (2011) photometric
estimates of galaxy redshifts in this work.

To define the QSO environment, we first
determined suitable values of r$_p$ and $\Delta V$ parameters by estimating
average galaxy surface density profiles and overdensities around the QSOs.
We selected galaxies around the QSOs 
with r${_p}$ from 0 to  500 kpc and
with $\Delta V $ from 0 to 12000 \kms.  
We used $z_Q$, the QSO spectroscopic redshift from SDSSQ--DR7
and $z_g$, the galaxy photometric redshift estimates from O' Mill et al.
(2011).  

\begin{figure}
     \includegraphics[width=80mm,height=80mm]{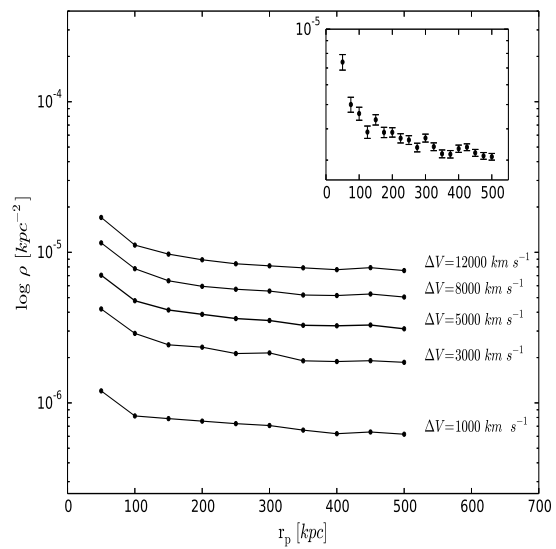}
   \caption{The density profile of galaxies in logarithmic scale obtained 
by fixing $\Delta V$ at different values and averaging over all studied 
QSOs. Right-upper panel shows the
density profile for $\Delta V$ = 5000 \kms including uncertainties. }
              \label{f1}%
    \end{figure}

For a given QSO, we obtained the galaxy 
surface density by counting the number of galaxies within concentric 
annuli of a 50  kpc  radius around the QSO fixing
the $\Delta V$ value. Subsequently, all the QSOs in the sample were
stacked to obtain the average profile.
Figure~\ref{f1} shows the density profile of galaxies 
  for five different $\Delta V$ values.  The profile obtained for
  $\Delta V = 5000$ \kms~is highlighted in the figure and it is also reproduced in
  the upper-right corner 
  in r$_p$ bins of 25  kpc.
  The density profiles have, in general,
  the same behavior: the density decreases down to a critical radius of $\sim$
  350 kpc and then it
  is almost constant as expected for a homogeneous background.  Comparable
  critical distances were also used in studying the effects
  of galaxy interactions on star formation in Nikolic, Cullen \& Alexander
  (2004).  
  Power-law fits of the density profiles were obtained for each $\Delta V$
  value.  A lower slope  is an indication of contamination and/or projection
  effects.  We found that the steepest profiles are obtained for $\Delta V $
  in the range 4000 to 6000 \kms.  
  
\begin{figure}
        \includegraphics[width=80mm,height=80mm]{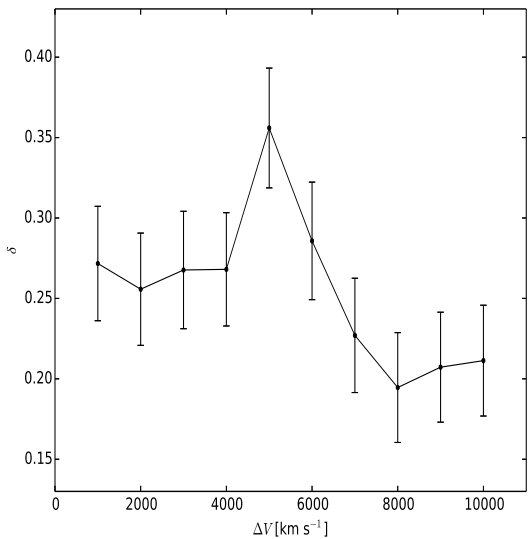}
        \caption{Galaxy overdensities obtained by fixing r$_p$ = 350  kpc and
          averaging over all studied 
QSOs. Error bars are uncertainties within each $\Delta V$ bin.}
                 \label{f1.1}%
   \end{figure}

Therefore, we counted galaxies (N$_{gal}$) within r$_p$ = 350 kpc to estimate 
galaxy overdensities centered in a given QSO and normalized it to the
expected number ($N_{mean}$) from the mean background density, $\delta = N_{gal} / N_{mean} - 1$. 
This mean 
  background density
was calculated taking into account
the densities obtained from random centers at the QSO redshifts with
r$_p$ up to 650  kpc.
 The overdensities were then calculated by counting the number of 
galaxies within the fixed
r$_p$ for different $\Delta V$ values 
and the final overdensity profile was averaged for all QSOs
in the sample.  The 
errors are the standard deviation in each $\Delta V$ bin.  
The average overdensity within 350 kpc as a function of $\Delta V$
is shown in Fig.~\ref{f1.1}.  This overdensity reaches a maximum 
value of
$\delta$ = 0.356 $\pm$ 0.037 at  $\Delta V $ = 5000 \kms~, in agreement
with the $\Delta V$ value providing the steepest density profile.
We notice that in spite of the individual error bars in the photometric
redshifts ($\sim$ 7000 km/s), the average
overdensity across the
QSO sample for each $\Delta V$ bin substantially reduces the uncertainties.

   Serber et al.  (2006) studied the small-scale environment of QSOs using
   only projected separations.  They showed 
   that the enhanced mean overdensity around the QSOs affects the
   100  kpc closest region.  The local density excess
   of neighboring galaxies within 100 to 500  kpc may contribute
   to trigger the QSO activity through mergers and interactions.  
   According to our analysis, r$_p$ of 350  kpc and
     $\Delta V$ of 5000 \kms~can be considered as reliable upper limits
     to define the QSO sub-samples. 
     To better establish the QSO interaction with neighbor galaxies, we defined
     r$_p$ = 140  kpc and $\Delta V = $ 5000 \kms~as a
   good compromise to be used in the definition of the QSO environment. 

\subsection{The QSO sub-samples}

Taking into account the sample of QSOs and their surrounding galaxies, we
defined three QSO sub-samples.  The extreme case was
defined as the strong QSO--galaxy interaction, the Sint sub-sample
with at least one galaxy
with 
projected separation $< 70 $ kpc and relative radial 
velocity  $<$ 5000 \kms.  A moderate case representing the weak
QSO--galaxy interaction, the Wint sub-sample corresponds to QSOs with a
companion galaxy with
$r_p $ between 70 and
140  kpc and $\Delta V$ $<$
5000 \kms.  Finally, for isolated QSOs, the Iso sub-sample corresponds to
objects without
companion galaxies
 with projected separations between 140 and 500  kpc and 5000 $ < \Delta V \le $ 12000 \kms. 

\begin{figure*}
   \includegraphics[width=70mm,height=70mm]{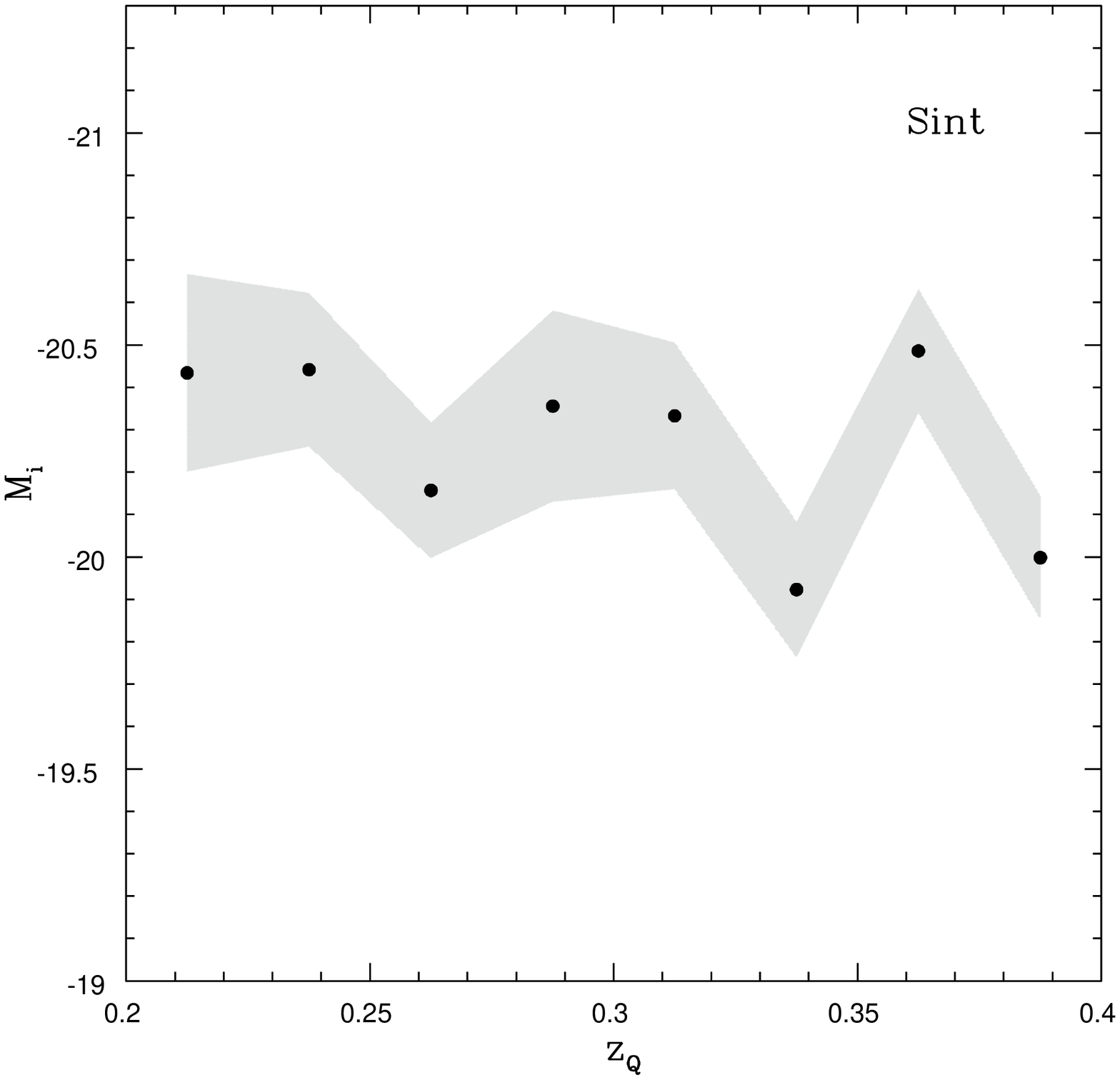}
     \includegraphics[width=70mm,height=70mm]{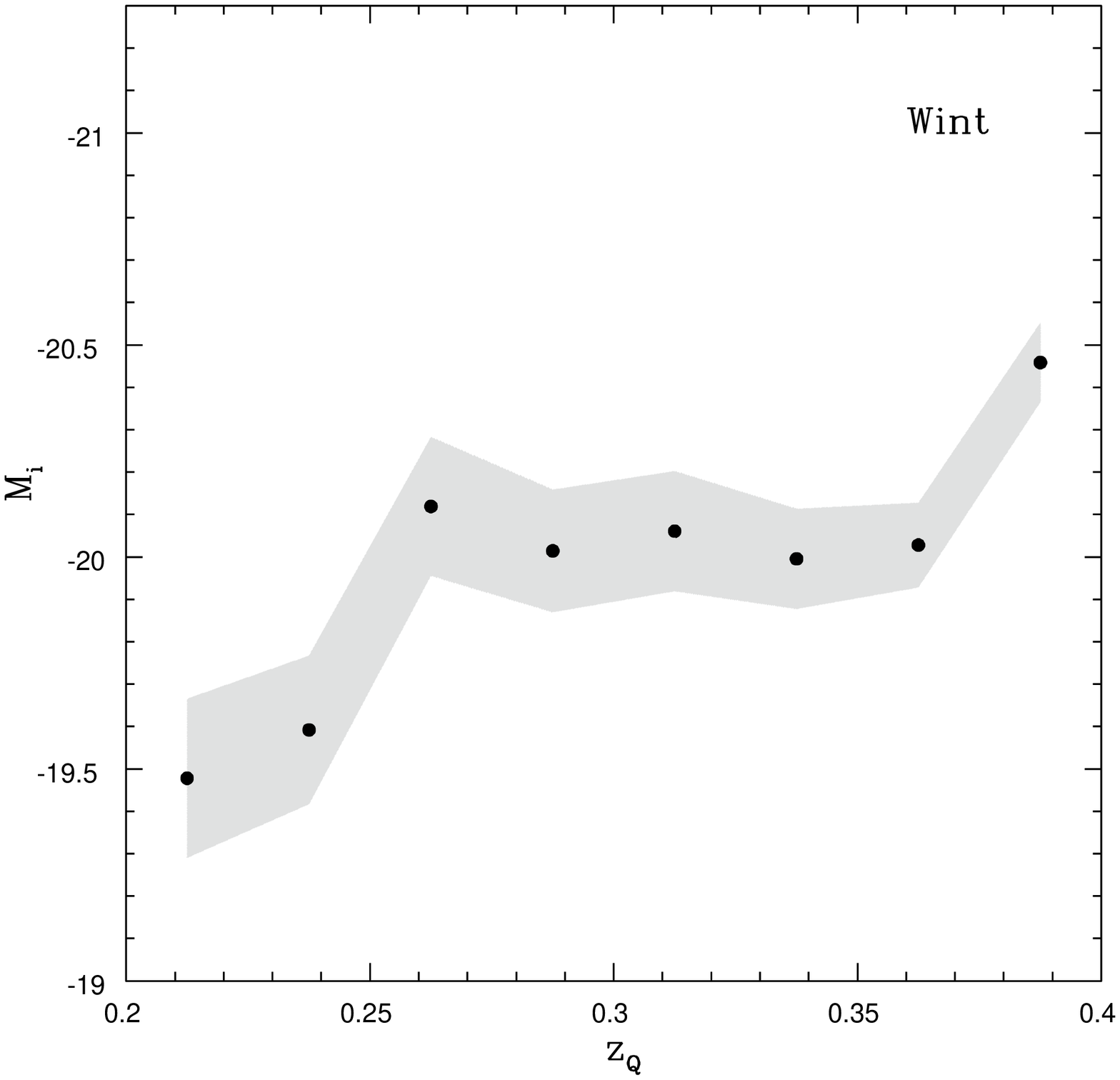}
     \caption{Median absolute magnitudes of the galaxies relative to
       the Sint and Wint
       sub-samples in redshift bins.  }
       \label{Estadistica}
\end{figure*}

Table~\ref{table1} shows a summary of the three defined QSO sub-samples 
with the 
adopted criteria and the number of objects. According to our criteria, 75\% 
of the QSOs  are isolated objects.  The Sint
sub-sample
includes the brightest QSOs with mean absolute magnitude differences of 0.1 and
0.6 mag with respect to the Wint and Iso sub-samples.
The QSO redshift distributions
are similar for the three sub-samples.
Figure~\ref{Estadistica} shows median absolute magnitudes 
of neighbor galaxies in redshift bins of 0.025 for
Sint (left) and Wint (right panel) sub-samples.
It can be seen that the neighbor galaxies of the Sint sub-sample are
generally brighter than those of the Wint sub-sample.  The limiting
magnitude given by O' Mill et al. (2011)
is fainter than the absolute
magnitudes of the galaxies in both sub-samples, except at z $\ge$ 0.35,
when they become comparable.  
  
        \begin{table*}
        \center
        \caption[]{The three defined QSO sub-samples.}
        \label{table1}
        \begin{tabular}{l c c r}
        \hline\hline 
            \noalign{\smallskip}
        Sub-Sample  & r$_p$ & $\Delta V$ & Number of QSOs  \\
          \noalign{\smallskip}
            \hline
            \noalign{\smallskip}
        Sint &  r$_p \leq $ 70  kpc & $\Delta V \leq $ 5000 km~s$^{-1}$  & 389 \\
        Wint  & 70  kpc $ < r_p \leq $ 140 kpc & $\Delta V \leq $ 5000 km~s$^{-1}$ &  751 \\ 
        Iso   & r$_p >$ 140  kpc & $\Delta V >$ 5000 km~s$^{-1}$ & 3523 \\

          \noalign{\smallskip}
            \hline
         \end{tabular}
   \end{table*}

        \section{Spectral analysis}

        \subsection{Statistical analysis}

From the SpecLineAll table of the CasJobs database, we extracted 
the EWs of the relevant spectral lines and their associated errors for
the studied QSOs.  We performed a visual inspection of the spectra to check
their quality.   We found some lines: H$_\epsilon$; [SII]$\lambda4072$;
G band $\lambda 4306$; [OI]$\lambda6366$ and
[SII]$\lambda\lambda6717,6731$ showing in general
low signal to noise ratios (S/Ns) and they were confused with the continuum.  The lines 
[NeV]$\lambda3346$,
H$_\eta$, HeI and 
[OI]$\lambda6302$, were only observed in certain QSO spectra (less than 20\%),
implying
few measurements with high uncertainties.  The line MgII $\lambda2798$ has,
in general, a good S/N, but it was
observed only for QSOs with z $>$ 0.35.  None of these lines were included in the following analysis.
For our study, the most important 
emission lines showing a better S/N were:  [NeV]$\lambda3426$; 
[OII]$\lambda\lambda3727,3730$;
H$_\delta$; H$_\gamma$; [OIII]$\lambda4363$; 
H$_\beta$; [OIII]$\lambda\lambda4959,5007$;
MgI$\lambda5177$; NaI$\lambda5896$; 
[NII]$\lambda6548$; 
H$_\alpha$ and [NII]$\lambda6583$. 

\begin{figure*}
        \centering
        \includegraphics[width=80mm,height=80mm]{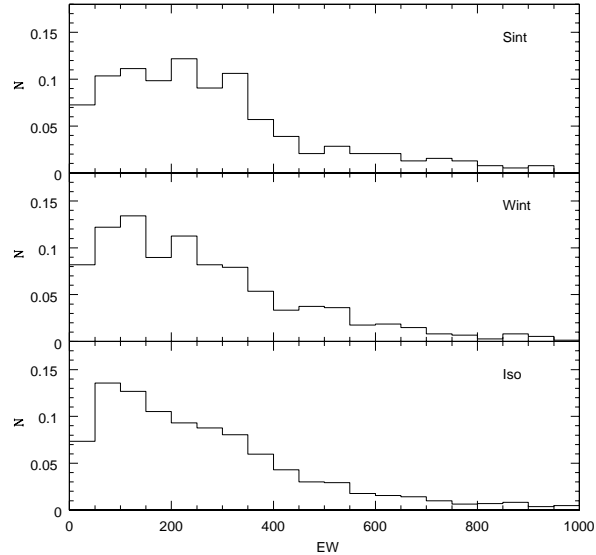}
        \caption{Normalized H$_\alpha$ EW distribution for the three defined 
QSO sub-samples.}
                \label{f2}
        \end{figure*}

Figure~\ref{f2} shows
the normalized EW distribution for the H$_\alpha$ emission lines 
for the Sint, Wint and Iso sub-samples, where it can be
observed that the distribution for the strong QSO-galaxy interaction 
sub-sample is almost constant for EW values from 50 to 350 \AA.
The distribution for the Wint sub-sample has a peak for EWs ranging from 50 to
150 \AA.  The same behavior is observed for the isolated QSOs between 50 and 100 \AA.  To avoid large 
outliers, which roughly 
represents 4$\sigma$ of the EW distribution, we only used spectral 
lines in our analysis when the ratio between 
the EW and the associated error was less than 0.3.

\begin{figure*}
  \centering
  \includegraphics[width=80mm,height=80mm]{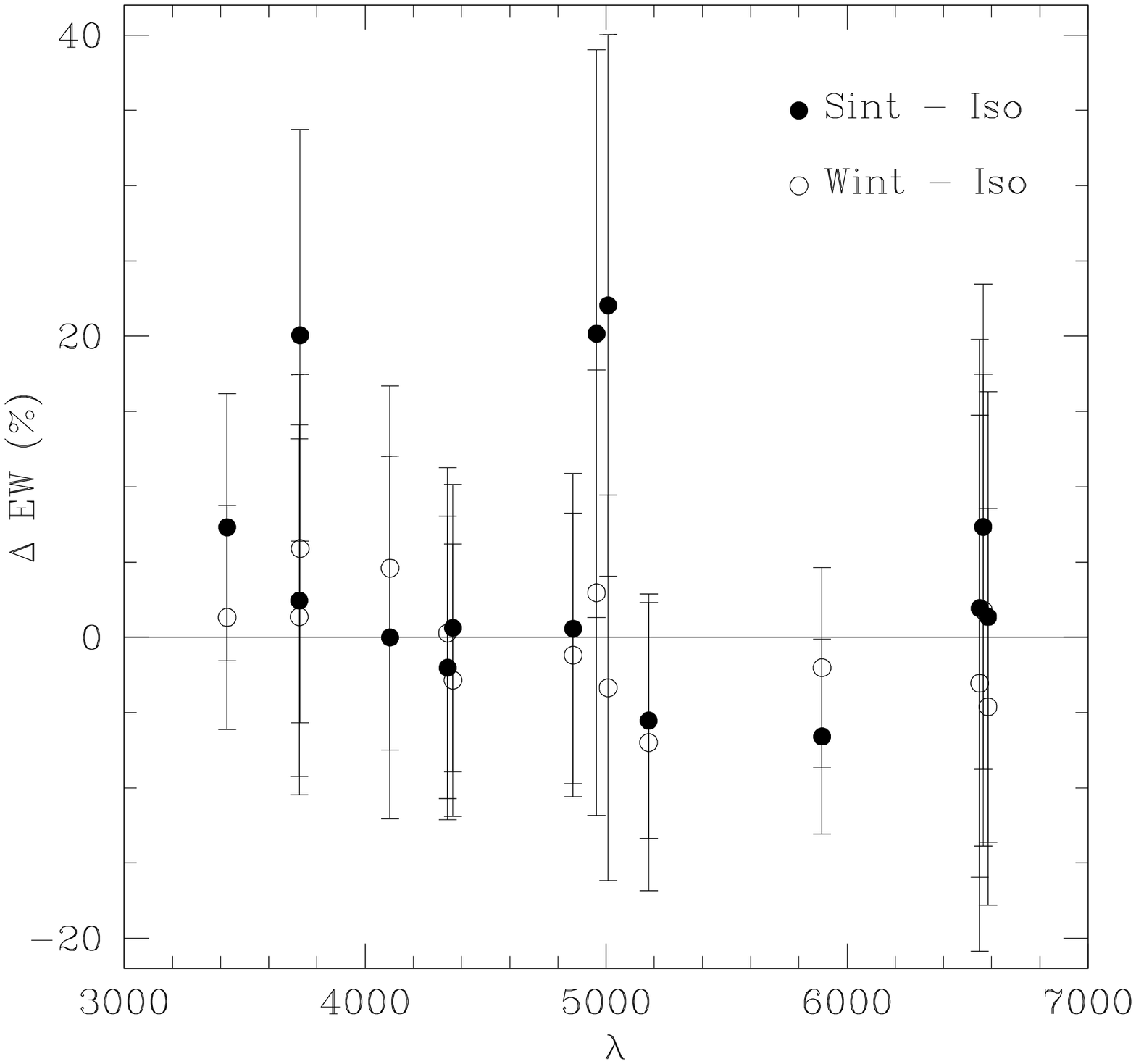}
\caption{The EW differences in percentage for the strong and weak QSO 
interactions with
galaxies relative to isolated QSOs. }
\label{f4}
\end{figure*}

To investigate how the interaction with neighbor galaxies
affects the QSOs, for each sub-sample 
we obtained median
EWs and uncertainties using bootstrapping 
of the most important spectral lines listed above.
Table~\ref{table2} shows the median
EW values and errors for the 14 studied spectral lines together with the
number of EW measurements for the
three different sub-samples. 
Figure~\ref{f4} displays the differences in percentage 
between 
the EW median values obtained for the strong and weak QSO-galaxy 
interactions with those for isolated QSOs. Five lines,
$[NeV] \lambda 3426$, [OII]$\lambda3730$,
[OIII]$\lambda\lambda4959,5007$ and 
H$_\alpha$  of the
QSOs in the Sint sub-sample, show larger median EW values when compared with
isolated QSOs.  As shown in the table, the number of measurements 
for $[NeV] \lambda 3426$ and [OII]$\lambda3730$ are about
half the number of measurements of the other three lines.  Also, in the case of
the [OII]$\lambda\lambda3727,3730$ lines, it is difficult
to separate the two lines with the available spectra.    They 
are not considered in the further analysis of the lines. For H$_\alpha$, the difference is around 7\% and for the [OIII]$\lambda\lambda4959,5007$ lines this difference is  more than
20\%.   The EW values for the QSOs in the Wint sub-sample
are smaller than 5\% with the exception of [OII]$\lambda3730$.  We argue that
the observed difference in the three most important
lines is related to the QSO interaction with
neighboring galaxies.

\begin{table*}
\center
\caption{The median EWs and uncertainties for the most relevant spectral lines 
in the three QSO sub-samples.}
\begin{tabular}{c r r r r r r }
\hline\hline
Spectral line & Sint & & Wint & & Iso & \\
\hline
& & & & & &\\
$[NeV] \lambda 3426$ & 3.67 $\pm$ 0.23 & 173 & 3.47 $\pm$ 0.17 & 309 & 3.42 $\pm$ 0.19 & 1450 \\
$[OII] \lambda 3727$ & 4.09 $\pm$ 0.34 & 96 & 4.05 $\pm$ 0.35 & 162 & 4.00 $\pm$ 0.31 & 663 \\
$[OII] \lambda 3730$ & 4.93 $\pm$ 0.38 & 171 & 4.35 $\pm$ 0.30 & 284 & 4.11 $\pm$ 0.34 & 1353 \\
H$_\delta$ & 8.55 $\pm$ 0.82 & 201 & 8.94 $\pm$ 0.80 & 349 & 8.55 $\pm$ 0.62 & 1612 \\
H$_\gamma$ & 14.58 $\pm$ 1.06 & 312 & 14.93 $\pm$ 1.22 & 572 & 14.89 $\pm$ 1.09 & 2699 \\
& & & & & & \\
$[OIII] \lambda 4363$ & 4.45 $\pm$ 0.29 & 276 & 4.30 $\pm$ 0.27 & 513 & 4.42 $\pm$ 0.30 & 2436 \\
H$_\beta$ & 37.88 $\pm$ 2.75 & 381 & 37.21 $\pm$ 2.31 & 732 & 37.66 $\pm$ 2.72 & 3399 \\
$[OIII] \lambda 4959$ & 8.63 $\pm$ 1.02 & 337 & 7.39 $\pm$ 0.74 & 604 & 7.18 $\pm$ 0.74 & 2856 \\
$[OIII] \lambda 5007$ & 19.92 $\pm$ 2.23 & 386 & 15.77 $\pm$ 1.44 & 738 & 16.32 $\pm$ 1.57 & 3477 \\
$MgI \lambda 5177$ & 9.44 $\pm$ 0.52 & 163 & 9.29 $\pm$ 0.80 & 314 & 9.99 $\pm$ 0.62 & 1429 \\
& & & & & & \\
$NaI \lambda 5896$ & 4.34 $\pm$ 0.23 & 127 & 4.56 $\pm$ 0.24 & 226 & 4.65 $\pm$ 0.20 & 1074 \\
$[NII] \lambda 6548$ & 29.48 $\pm$ 3.63 & 376 & 28.04 $\pm$ 3.78 & 727 & 28.93 $\pm$ 3.61 & 3391 \\
H$_\alpha$ & 248.56 $\pm$ 24.33 & 386 & 235.73 $\pm$ 24.42 & 745 & 231.57 $\pm$ 26.35 & 3494 \\
$[NII] \lambda 6583$ & 26.85 $\pm$ 2.82 & 380 & 25.27 $\pm$ 2.31 & 716 & 26.49 $\pm$ 2.75 & 3363 \\
\hline 
\end{tabular}
\label{table2}
\end{table*}

To add further statistical significance to the
results, we applied the Kolmogorov-Smirnov (KS) test
to the distributions of these three spectral lines. We compared
the EW distributions of the Sint relative to the Iso sub-sample, finding that
for the [OIII]$\lambda\lambda4959,5007$ lines they are different
distributions. 
For the Wint relative to Iso sub-sample, the results
show that the distributions are drawn from the same parent
population. With the H$_\alpha$ line, the results show the same parent population
in both sub-samples relative to the isolated QSOs.

We also used bootstrap in the EW distribution for these three  lines
 to provide robust uncertainty bounds for our median estimates. 
 We computed the bootstrapped 95\% confidence intervals (CIs) around the median
 EW values.
 Table~\ref{table3} shows the CIs for the three studied spectral lines for the
 different QSO sub-samples.  The results show higher median EWs within
 wider CIs for the
 Sint sub-sample in comparison with the other two sub-samples.  The most
 evident difference is related to the H$_\alpha$ line.

\begin{table*}
\center
\caption{Confidence intervals for EW distributions in the three QSO sub-samples.}
\begin{tabular}{c c c c }
\hline\hline
Spectral line & Sint & Wint & Iso \\
\hline
& & & \\
$[OIII] \lambda4959$ & $[7.39, 9.70]$ & $[6.80, 7.94]$ & $[6.85, 7.40]$ \\
$[OIII] \lambda5007$ & $ [17.73, 22.09]$ & $[14.79, 16.61]$ & $[15.58,16.64]$ \\
H$_\alpha$ & $ [228.40, 273.27]$ & $[215.16, 247.65]$ & $[222.67, 241.10]$ \\

\hline 
\end{tabular}
\label{table3}
\end{table*}

We also restricted the definition of the strong QSO-galaxy interaction 
by using a smaller $\Delta V $ of 3000 \kms, implying
235, 486 and 3942 QSOs in the new Sint, Wint, and Iso sub-samples, 
respectively.  We obtained median EW values for the 
[O{\scriptsize III}]$\lambda4959,5007$ and H$_\alpha$ lines that showed 
similar results to previous percentage differences.  The observed increments in the
[O{\scriptsize III}] EW could  be related to 
strong AGN activity (Kauffmann et al.
2003) induced by galaxy interactions.  

The results obtained for the H$_\alpha$ line for QSOs with strong
galaxy interactions is 
consistent with some studies of normal galaxy pairs.
Bushouse, Werner \& Lamb (1988) and
Donzelli \& Pastoriza (1997) have shown that the nuclear star 
formation increases in galaxy
interactions and mergers.  
In these sub-samples, less than 10\% of
the pairs are low-luminosity AGNs (LINER and Seyfert II galaxies).
Barton Gillespie, Geller \& Kenyon (2003) 
studied recent
star formation in galaxy pairs, and demonstrated 
that the H$_\alpha$ EWs were strongly correlated with the inverse of the
pair spatial separation or velocity difference. Using stellar
population synthesis models, they obtained a merger
scenario in which close interaction triggers starburst and
increases the EWs.

\subsection{Spectral decomposition}

In this subsection, we explore the importance of the host galaxy and the role of the environment on the QSO 
activity.  We analyzed the SDSS spectra of the QSOs  with z $<$ 0.31 with
a well defined H$_{\alpha}$ emission line.   In this redshift range, we obtained a sub-sample of
100 QSOs in the restricted Sint sub-sample and also 100 randomly selected QSOs in the Iso sub-sample, which represents about 10\% of the total.

\begin{figure*}
 \centering 
 \includegraphics[width=60mm,height=60mm]{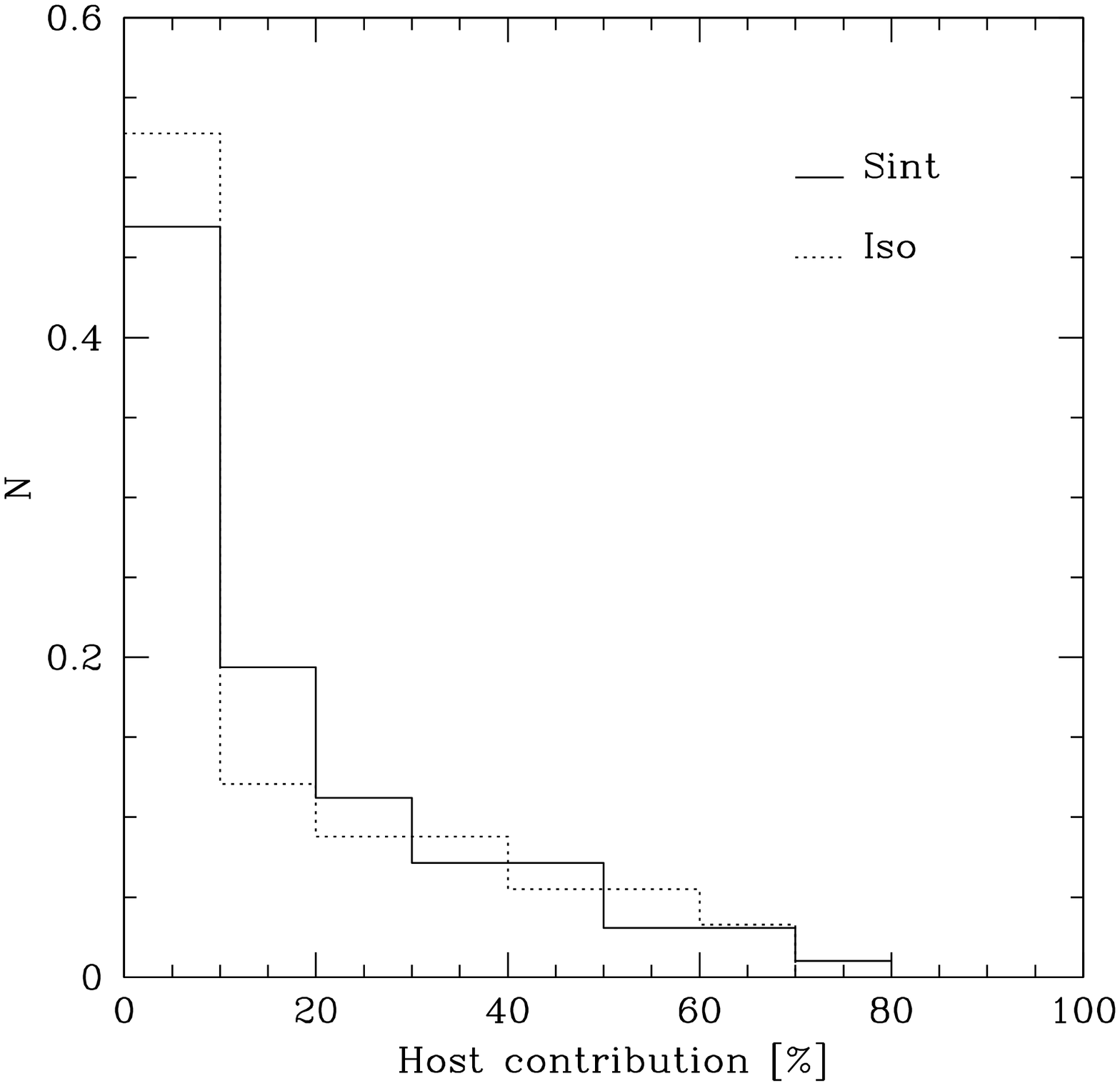}
   \includegraphics[width=60mm,height=60mm]{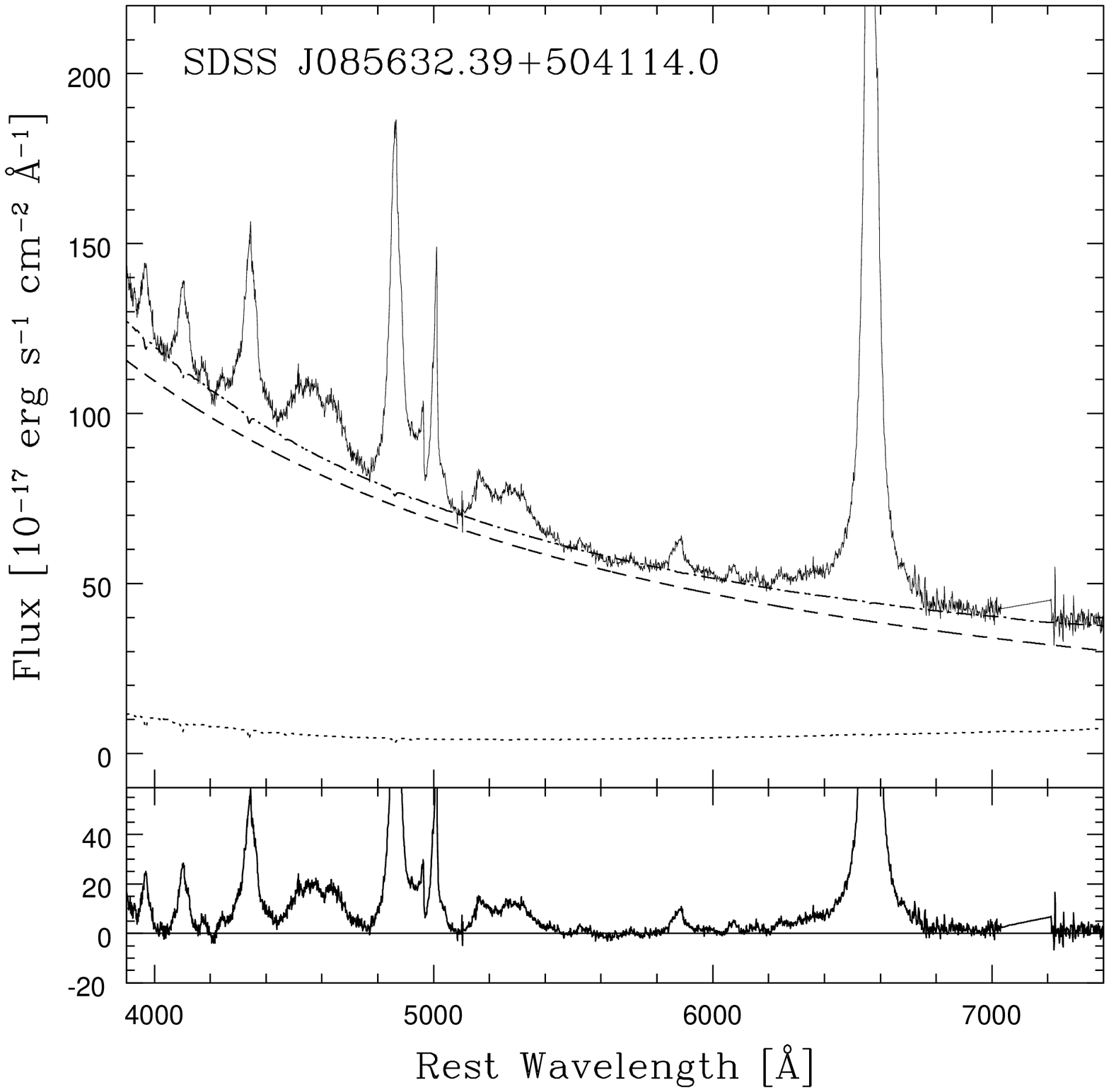}
   \caption{
     Left panel:  Normalized distributions of the host contribution for the Sint and Iso
    sub-samples represented by solid and dashed lines, respectively.
    Right panel: Example of the STARLIGHT fitting of a QSO
    spectrum.   The solid line represents the observed
    spectrum; dot - short dash line,
    the synthethic spectrum; short dash line, the AGN contribution in a
    power-law form, and dotted line, the host contribution (upper panel).  The
    bottom panel shows the residuals.}
\label{f4.1}
\end{figure*}

  First, we estimated the contribution of the host galaxy to the total emission.  We applied the STARLIGHT code
  (Cid Fernandes et al.
  2005, Mateus et al. 2006) that combines empirical
population synthesis with  evolutionary synthesis
models. We used a base consisting of 80 single stellar population templates with different ages and metallicities 
  (Bruzual \& Charlot 2003) and six power-law components (e.g., Cardoso et al. 2017).  
Considering these two QSO sub-samples, we found that the contribution of the
host galaxy to the total emission  is 11\% in median  for the restrictive
Sint sub-sample and 8\% for the isolated QSOs.
This contribution is in agreement with
previous works, for example, at lower redshifts;  Younes et al. (2010)
studying the galaxy NGC 4278 at z = 0.002.  At higher redshifts,
the stellar contribution was estimated to be around 5 to 25\% (Dey et al. 2008 and
Donley et al. 2010) for a sample of galaxies at z = 2.   
The left panel of Fig.~\ref{f4.1} shows the normalized distribution of 
the host contribution to the total emission
for the restricted Sint 
(solid line) and Iso (dot line) sub-samples.
We estimated that 80\% of the QSOs have host contributions smaller than 30\%
and the lowest contributions are found for the brightest QSOs in the sample
(Fig.~\ref{histoM}).  The right panel of the figure shows as an example, the STARLIGHT results of a
QSO spectrum with 9\% of
the host contribution to the total emission.  In the upper panel, we show
the observed
    spectrum (solid line); 
    the synthetic spectrum (dot - short dash line) obtained from 
    the AGN contribution in a 
    power-law form (short dash line); and the host contribution (dotted line).
 The bottom
    panel shows  the residuals
    between the observed and synthethic spectra, which correspond to the emission spectrum of
    the ionized gas. 

The H$_{\alpha}$ emission line can be decomposed into narrow and
broad components, and these might be affected by different physical processes
(Antonucci 1993) originated in different regions (the NLR and BLR).
An increase in the narrow emission component may indicate an important 
contribution from  enhanced star formation, while an increment in the broad
component would be related to the AGN itself.  However, from the analysis of 
median EW values alone, it is not
possible to disentangle the different component
contributions. Therefore, in the present study, we assumed two components for 
the  H$_{\alpha}$ emission 
line profiles by adopting a combination of
Gaussian profiles.  The LINER routine 
(Pogge \& Owen 1993) based on a $\chi^2$ minimization algorithm was used to fit 
the narrow and broad components for the H$_{\alpha}$ profile 
with some assumptions (e.g., Schmidt et al. 2016).
One of these is that [NII]$\lambda6548$ and 
[NII]$\lambda6583$ are considered to be of equal FWHM, because both lines 
are emitted in the same region. The second one is that the flux ratio of the 
two [NII] components is equal to their theoretical value of 1:3. For some 
objects, it was also necessary to impose a third constraint for these lines: 
that their wavelength separation is equal to their theoretical value (36 \AA).
In most of the cases, we obtained FWHM values of approximately 
400 \kms~and 3000 \kms~ for the narrow and broad components of the H$_{\alpha}$ line, respectively, in agreement with previous results (e.g., Marziani et al. 2010,
Coatman et al. 2017).

\begin{figure*}
 \centering 
  \includegraphics[width=80mm,height=80mm]{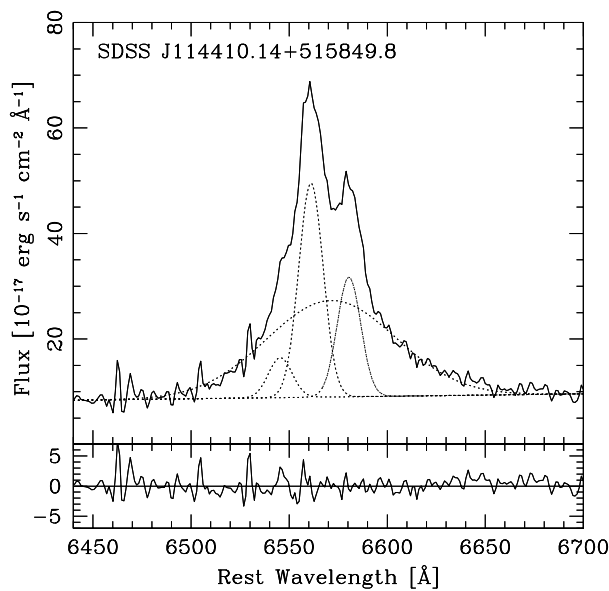}
\caption{Fit of a typical H$_{\alpha}$ emission line with the 
[NII]$\lambda\lambda6548,6583$ lines.  The thick black line shows the observed 
spectrum and the dotted lines display the Gaussian profile decomposition 
together with, the [NII]$\lambda\lambda6548,6583$ lines.
 The bottom panel shows the residuals.}
\label{fline}
\end{figure*}

Figure~\ref{fline} shows a typical fit of the H$_{\alpha}$ emission line 
together with the [NII]$\lambda\lambda6548,6583$ lines, and also the corresponding
residuals.  The observed emission spectrum is represented by a thick black
line, with the different Gaussian components, including the H$_{\alpha}$ broad 
and narrow lines and the [NII], represented by dotted lines.

\begin{figure*}
  \includegraphics[width=80mm,height=80mm]{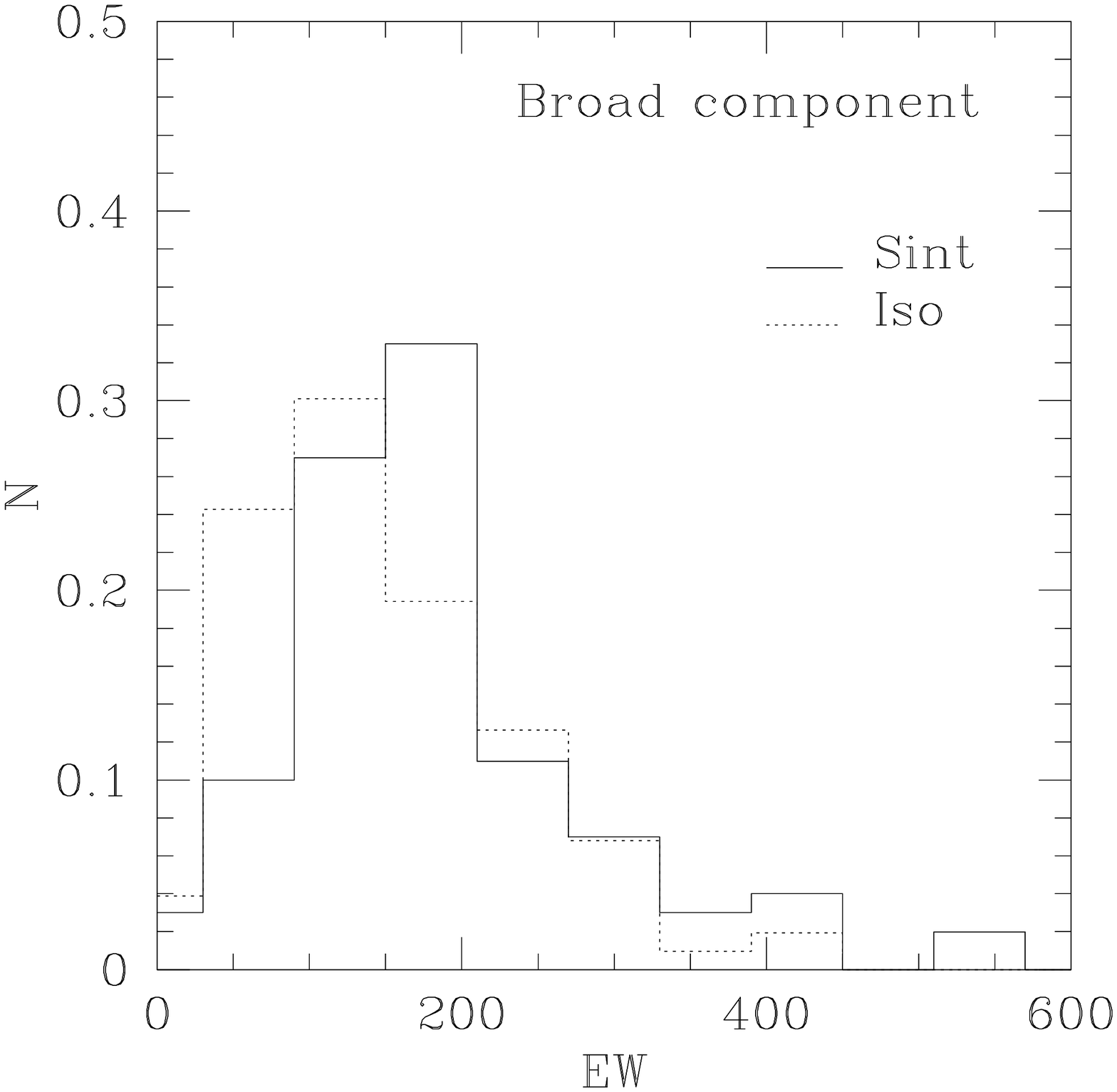}
  \includegraphics[width=80mm,height=80mm]{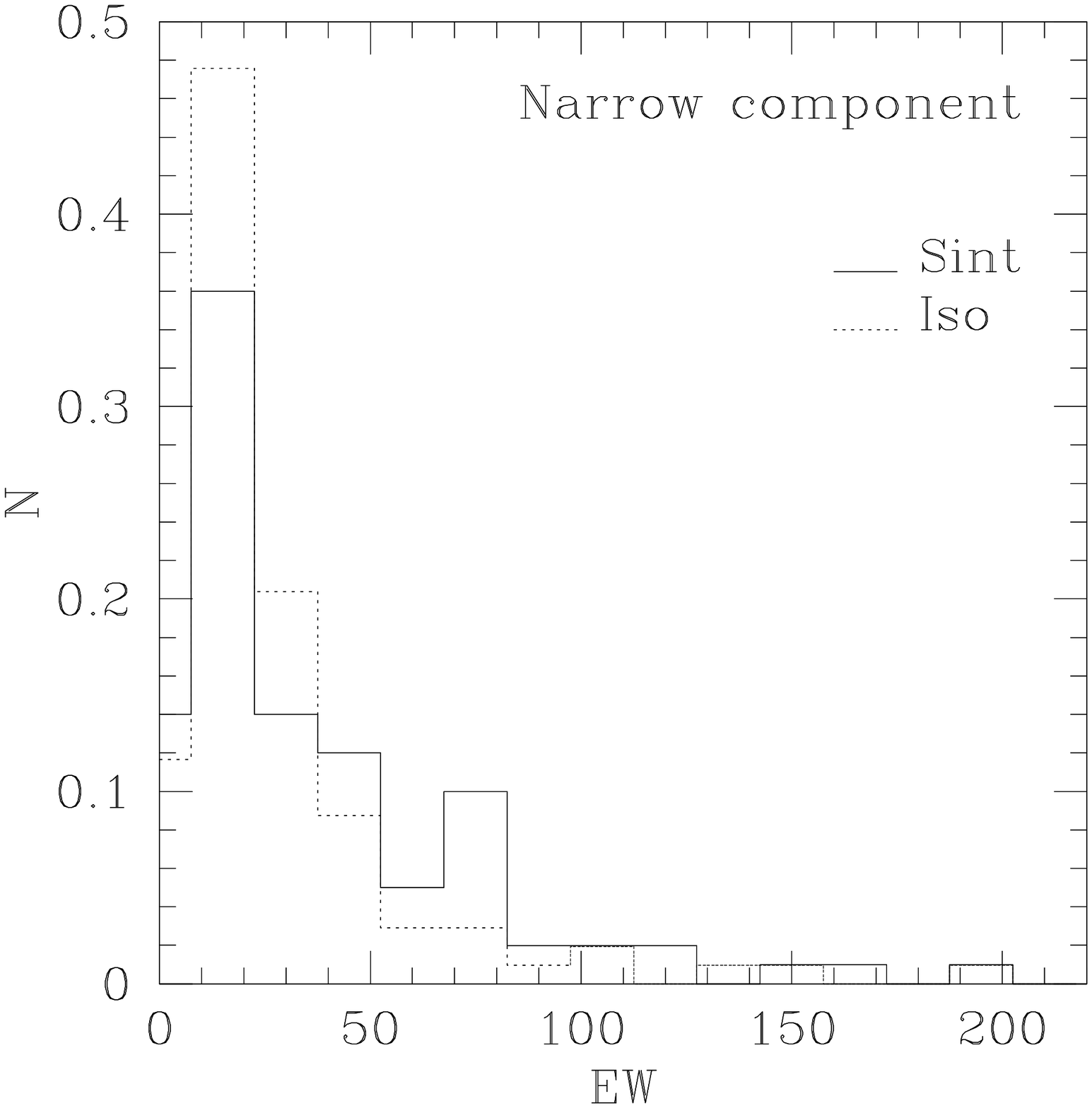}
\caption{Normalized equivalent width distributions of the H$_{\alpha}$ emission 
line for the broad 
(left panel) and narrow (right panel) components.  The solid and dotted
lines correspond to strong interacting and isolated QSOs in the 
restricted sub-samples, respectively.}
\label{fcomp}
\end{figure*}

Figure~\ref{fcomp} shows the results of the decomposition in broad (left
panel) and
narrow  (right panel) H$_{\alpha}$ components of the two QSO sub-samples.
The solid (dot) lines represent
the EW distributions for the QSOs from the restricted
Sint (Iso) sub-samples.  For the analyzed QSOs that had a strong interaction with neighboring 
galaxies, the median values were  22.60 $\pm$ 4.59~\AA~for the narrow 
component and 161.40 $\pm$ 12.92~\AA~for the broad one. 
For the isolated QSOs, we obtained a median value
of 18.71 $\pm$ 3.87~\AA\  for the narrow component,~and 135.52 $\pm$ 10.42~\AA\  for the broad one.
Within the uncertainties, the distribution of the narrow component was similar 
for QSOs with strong 
interaction and in isolation.  However, the broad components of interacting 
QSOs were 20\% wider than in isolated QSOs. This result might show that the 
connection between the galaxy interactions and QSO activity was via the central 
regions.  Considering that the broad component of H$_{\alpha}$ is originated at the BLR, the emission lines which could be
contaminated are the narrow lines. According to the synthesis results, this contamination is around 10\% in median.

Finally, we applied the KS test to the EW measurements of the narrow and broad components
obtained from the decomposition of the H$_{\alpha}$ emission line to the
Sint and Iso sub-samples.  The results show that the broad component
distributions are two different
populations while the narrow components 
are similar distributions.

\section{Summary}

We defined a QSO sample using 
the SDSS Quasar catalog derived from the SDSS-DR7 (Schneider et al. 2010), 
which was 
comprised of 4663 QSOs in the redshift range of 0.20 $ < z \leq $ 0.40.  In order to study the effects in the QSO spectra produced by 
the environment, we defined three different QSO sub-samples.  We selected
neighboring
galaxies relative to the QSOs using the projected separations and radial
velocity
differences. Using a projected separation of 140 kpc and 
$\Delta V$ = 5000 \kms,
 three QSO sub-samples were defined representing the strong and
weak interactions with galaxies, and isolated QSOs. These QSO sub-samples
were also restricted
using a smaller $\Delta V $ value of 3000 \kms~to perform statistical
and spectral analyses.  

From the study of the EW medians of the most relevant spectral lines, we
found that the [OIII]$\lambda\lambda4959,5007$ lines show values larger than 20\%
for QSOs in the Sint sub-sample when compared with isolated QSOs, and about 7\% for the 
H$_\alpha$ line.  The changes in EWs were, in general, comparable between the weak QSO-galaxy
interactions and  isolated QSOs.   For these three lines, we also performed
the KS test and obtained confidence intervals for the EW medians.  For the
[OIII]$\lambda\lambda4959,5007$ lines, the EW distributions
of the Sint  are different from those of the Iso sub-sample.  For the Wint relative to Iso sub-sample, the results
show that the distributions are drawn from the same parent
population.  For the  H$_\alpha$ line, the CIs have higher median EWs within
 wider intervals for the
 Sint sub-sample in comparison with the other two sub-samples.  
 These results show that only some
line EWs of QSOs are marginally influenced by environment.
Similar results were obtained when
restricting the sub-samples with $\Delta V = $ 3000 \kms.  None of these results are affected by the emission of the host galaxy, which contributes
about 10\% in median to the total emission.  

To gain a better understanding of the connection between 
the increment of the H$_{\alpha}$ 
emission line and the  effect produced by the QSO-galaxy interactions, we  
performed broad and narrow line decomposition.  Throughout this decomposition, 
for the restricted sub-samples, the narrow component remained constant for QSOs with strong galaxy interactions and isolated QSOs.  
In contrast, the broad component revealed
an increase for the QSOs with strong galaxy interactions. 
Our results suggest that the QSO central regions (BLR and inner NLR)
could be mostly affected by
the galaxy interaction.  This conclusion is derived from the increase of the
broad component of H$_{\alpha}$ for the Sint sub-sample and the lack of variation in the narrow component of H$_{\alpha}$
and the
$[NII] \lambda 6583$ lines,  both originated in the NLR.  
This  is also reinforced by the changes in the  [OIII]$\lambda\lambda4959,5007$ lines, which are originated mainly in the inner parts of the NLR due to
their higher degree of ionization.  

Our findings suggest that QSO interactions with
neighboring galaxies may affect  AGN
activity as evidenced by the EW values of some emission lines.  A scenario where the accretion rate of massive black holes
is triggered by galaxy interactions as suggested by
Di Matteo, Springel \& Hernquist (2005) simulations could provide important clues about the joint
evolution of QSOs and galaxies.

\begin{acknowledgements}

  We would like to thank the anonymous referee for the suggestions to improve the manuscript.
  This work was partially supported by Consejo de Investigaciones 
Cient\'ificas y T\'ecnicas (CONICET) and Secretar\'ia de Ciencia y T\'ecnica de la Universidad Nacional de C\'ordoba (SecyT).  

SDSS-III is managed by the Astrophysical Research Consortium of the Participating Institutions of the SDSS-III Collaboration including the University of Arizona, the Brazilian Participation Group, Brookhaven National Laboratory, University of Cambridge, University of Florida, the French Participation Group, the German Participation Group, the Instituto de Astrof\'isica de Canarias, the Michigan State/Notre Dame/JINA Participation Group, Johns Hopkins University, Lawrence Berkeley National Laboratory, Max Planck Institute for Astrophysics, New Mexico State University, New York University, Ohio State University, Pennsylvania State University, University of Portsmouth, Princeton University, the Spanish Participation Group, University of Tokyo, University of Utah, Vanderbilt University, University of Virginia, University of Washington, and Yale University.
\end{acknowledgements}


\begin{thebibliography}{}

\bibitem[Abazajian et al.(2004)]{2004AJ....128..502A} Abazajian, K., 
Adelman-McCarthy, J.~K., Ag{\"u}eros, M.~A., et al.\ 2004, \aj, 128, 502 

\bibitem[Abazajian et al.(2009)]{2009ApJS..182..543A} Abazajian, K.~N., 
Adelman-McCarthy, J.~K., Ag{\"u}eros, M.~A., et al.\ 2009, \apjs, 182, 543 

\bibitem[Antonucci(1993)]{1993ARA&A..31..473A} Antonucci, R.\ 1993, \araa, 31, 473 

\bibitem[Bahcall et al.(1997)]{1997quho.conf...37B} Bahcall, J.~N., 
Kirhakos, S., \& Schneider, D.~P.\ 1997, Quasar Hosts, 37 

\bibitem[Barton Gillespie et al.(2003)]{2003ApJ...582..668B} Barton 
  Gillespie, E., Geller, M.~J., \& Kenyon, S.~J.\ 2003, \apj, 582, 668


\bibitem[Bruzual \& Charlot(2003)]{2003MNRAS.344.1000B} Bruzual, G., \& Charlot, S.\ 2003, \mnras, 344, 1000
  
\bibitem[Bushouse et al.(1988)]{1988ApJ...335...74B} Bushouse, H.~A., 
  Werner, M.~W., \& Lamb, S.~A.\ 1988, \apj, 335, 74

  \bibitem[Cardoso et al.(2017)]{2017A&A...604A..99C} Cardoso, L.~S.~M., Gomes, J.~M., \& Papaderos, P.\ 2017, \aap, 604, A99 

\bibitem[Carswell et 
  al.(1976)]{1976A&A....53..275C} Carswell, R.~F., Coleman, G., Williams, R.~E., \& Strittmatter, P.~A.\ 1976, \aap, 53, 275

  \bibitem[Cid Fernandes et al.(2005)]{2005MNRAS.358..363C} Cid Fernandes, R., Mateus, A., Sodr{\'e}, L., Stasi{\'n}ska, G., \& Gomes, J.~M.\ 2005, \mnras, 358, 363 


\bibitem[Coatman et al.(2017)]{2017AAS...22930203C} Coatman, L., Hewett, P.~C., Banerji, M., et al.\ 2017, American Astronomical Society Meeting Abstracts \#229, 229, 302.03
  
\bibitem[Collister 
\& Lahav(2004)]{2004PASP..116..345C} Collister, A.~A., \& Lahav, O.\ 2004, \pasp, 116, 345 

\bibitem[Connolly et al.(1995)]{1995AJ....110.2655C} Connolly, A.~J., 
  Csabai, I., Szalay, A.~S., et al.\ 1995, \aj, 110, 2655

  \bibitem[Dey et al.(2008)]{2008ApJ...677..943D} Dey, A., Soifer, B.~T., Desai, V., et al.\ 2008, \apj, 677, 943-956 

\bibitem[Di Matteo et al.(2005)]{2005gbha.conf..340D} Di Matteo, T., 
Springel, V., 
\& Hernquist, L.\ 2005, Growing Black Holes: Accretion in a Cosmological Context, 340

\bibitem[Donley et al.(2010)]{2010ApJ...719.1393D} Donley, J.~L., Rieke, G.~H., Alexander, D.~M., Egami, E., \& P{\'e}rez-Gonz{\'a}lez, P.~G.\ 2010, \apj, 719, 1393

\bibitem[Donzelli 
\& Pastoriza(1997)]{1997ApJS..111..181D} Donzelli, C.~J., \& Pastoriza, M.~G.\ 1997, \apjs, 111, 181 

\bibitem[Dunlop et al.(2003)]{2003MNRAS.340.1095D} Dunlop, J.~S., McLure, 
R.~J., Kukula, M.~J., et al.\ 2003, \mnras, 340, 1095 

\bibitem[Falomo et al.(2014)]{2014MNRAS.440..476F} Falomo, R., Bettoni, D., Karhunen, K., Kotilainen, J.~K., \& Uslenghi, M.\ 2014, \mnras, 440, 476 

\bibitem[Floyd et al.(2004)]{2004MNRAS.355..196F} Floyd, D.~J.~E., Kukula, 
M.~J., Dunlop, J.~S., et al.\ 2004, \mnras, 355, 196 


\bibitem[Heckman et al.(2004)]{2004ApJ...613..109H} Heckman, T.~M., 
Kauffmann, G., Brinchmann, J., et al.\ 2004, \apj, 613, 109 

\bibitem[Jahnke et al.(2004)]{2004MNRAS.352..399J} Jahnke, K., Kuhlbrodt, 
B., \& Wisotzki, L.\ 2004, \mnras, 352, 399 

\bibitem[Jogee(2006)]{2006LNP...693..143J} Jogee, S.\ 2006, Physics of 
Active Galactic Nuclei at all Scales, 693, 143 

\bibitem[Kauffmann et al.(2003)]{2003MNRAS.346.1055K} Kauffmann, G., 
Heckman, T.~M., Tremonti, C., et al.\ 2003, \mnras, 346, 1055


\bibitem[Lambas et al.(2003)]{2003MNRAS.346.1189L} Lambas, D.~G., Tissera, 
P.~B., Alonso, M.~S., \& Coldwell, G.\ 2003, \mnras, 346, 1189

\bibitem[Lynden-Bell(1969)]{1969Natur.223..690L} Lynden-Bell, D.\ 1969, 
  \nat, 223, 690

  \bibitem[Mateus et al.(2006)]{2006MNRAS.370..721M} Mateus, A., Sodr{\'e}, L., Cid Fernandes, R., et al.\ 2006, \mnras, 370, 721 

\bibitem[Marziani et al.(2010)]{2010MNRAS.409.1033M} Marziani, P., Sulentic, J.~W., Negrete, C.~A., et al.\ 2010, \mnras, 409, 1033
  
\bibitem[O' Mill et al.(2011)]{2011MNRAS.413.1395O} O' Mill, A.~L., 
Duplancic, F., Garc\'ia Lambas, D., 
\& Sodr\'e, L., Jr.\ 2011, \mnras, 413, 1395

\bibitem[Nikolic et al.(2004)]{2004MNRAS.355..874N} Nikolic, B., Cullen, H., \& Alexander, P.\ 2004, \mnras, 355, 874 

\bibitem[Osterbrock(1989)]{1989NYASA.571...99O} Osterbrock, D.~E.\ 1989, 
Annals of the New York Academy of Sciences, 571, 99 

\bibitem[P{\^a}ris et al.(2014)]{2014A&A...563A..54P} P{\^a}ris, I., Petitjean, P., Aubourg, {\'E}., et al.\ 2014, \aap, 563, A54

\bibitem[P{\^a}ris et al.(2017)]{2017A&A...597A..79P} P{\^a}ris, I., Petitjean, P., Ross, N.~P., et al.\ 2017, \aap, 597, A79 

\bibitem[Petrosian(1976)]{1976ApJ...209L...1P} Petrosian, V.\ 1976, \apjl, 
209, L1 

\bibitem[Pogge \& Owen(1993)]{} {Pogge}, R.~W. and {Owen}, J.~M.\ 1993,
  OSU Internal Report 93-01

  \bibitem[Scranton et al.(2002)]{2002ApJ...579...48S} Scranton, R., Johnston, D., Dodelson, S., et al.\ 2002, \apj, 579, 48 

\bibitem[Schmidt(1969)]{1969ARA&A...7..527S} Schmidt, M.\ 1969, \araa, 7, 527

 \bibitem[Schmidt et al.(2016)]{2016A&A...596A..95S} Schmidt, E.~O., Ferreiro, D., Vega Neme, L., \& Oio, G.~A.\ 2016, \aap, 596, A95

\bibitem[Schneider et al.(2010)]{2010AJ....139.2360S} Schneider, D.~P., 
Richards, G.~T., Hall, P.~B., et al.\ 2010, \aj, 139, 2360 

\bibitem[Serber et al.(2006)]{2006ApJ...643...68S} Serber, W., Bahcall, N., 
M\'enard, B., \& Richards, G.\ 2006, \apj, 643, 68 

\bibitem[Strauss et al.(2002)]{2002AJ....124.1810S} Strauss, M.~A., 
Weinberg, D.~H., Lupton, R.~H., et al.\ 2002, \aj, 124, 1810 

\bibitem[Strittmatter 
\& Williams(1976)]{1976ARA&A..14..307S} Strittmatter, P.~A., \& Williams, R.~E.\ 1976, \araa, 14, 307

\bibitem[Stoughton et al.(2002)]{2002AAS...20111412S} Stoughton, C., Lin, 
H., Yanny, B., et al.\ 2002, Bulletin of the American Astronomical Society, 
34, \#114.12 

\bibitem[Urry 
\& Padovani(1995)]{1995PASP..107..803U} Urry, C.~M., \& Padovani, P.\ 1995, \pasp, 107, 803 

\bibitem[York et al.(2000)]{2000AJ....120.1579Y} York, D.~G., Adelman, J., 
  Anderson, J.~E., Jr., et al.\ 2000, \aj, 120, 1579

  \bibitem[Younes et al.(2010)]{2010A&A...517A..33Y} Younes, G., Porquet, D., Sabra, B., et al.\ 2010, \aap, 517, A33 

\bibitem[Zel'dovich 
\& Novikov(1965)]{1965SPhD....9..834Z} Zel'dovich, Y.~B., \& Novikov, I.~D.\ 1965, Soviet Physics Doklady, 9, 834 


\end{thebibliography}
\end{document}